\RequirePackage{fix-cm}
\documentclass[smallcondensed]{svjour3}     
\smartqed  
\usepackage{graphicx}
\usepackage{amsmath}
\usepackage{comment}

\begin{document}

\title{Effects of interaction imbalance in a strongly repulsive one-dimensional Bose gas}


\author{R. E. Barfknecht \and A. Foerster \and N. T. Zinner}


\institute{R. E. Barfknecht \at{Department of Physics and Astronomy, Aarhus University, Ny Munkegade 120, Denmark \\ \email{rafael@phys.au.dk} \and Instituto de F{\'i}sica da UFRGS, Av. Bento Gon{\c c}alves 9500, Porto Alegre, RS, Brazil}
\\ \email{rafael.barfknecht@ufrgs.br}        
\and
A. Foerster \at Instituto de F{\'i}sica da UFRGS, Av. Bento Gon{\c c}alves 9500, Porto Alegre, RS, Brazil
\and
N. T. Zinner \at{Department of Physics and Astronomy, Aarhus University, Ny Munkegade 120, Denmark \and Aarhus Institute of Advanced Studies, Aarhus University, DK-8000 Aarhus C, Denmark}}

\date{Received: date / Accepted: date}

\maketitle

\begin{abstract}
We calculate the spatial distributions and the dynamics of a few-body two-component strongly interacting Bose gas confined to an effectively one-dimensional trapping potential. We describe the densities for each component in the trap for different interaction and population imbalances. We calculate the time evolution of the system and show that, for a certain ratio of interactions, the minority population travels through the system as an effective wave packet.
\end{abstract}

\section{Introduction}\label{intro}
The theoretical study of one-dimensional models has been pushed forward in recent years due to the possibility of constructing and measuring such systems in experiments with ultracold atoms. Examples of successful applications of these experimental techniques are the realization of the Tonks-Girardeau gas of infinitely repulsive bosons \cite{weiss2,paredes} and the observation of a cold atomic system that does not exhibit thermalization \cite{weiss1}. Furthermore, by manipulating the trapping potential, systems with only a few atoms have been constructed, and their properties have been measured \cite{jochim1,jochim2}. As common features, these experiments exhibit the control over the interactions between the atoms, often exploring the strongly correlated regime. 

To keep up with these advances, one specific theoretical approach has gained attention: a strongly interaction system of atoms with two or more internal degrees of freedom can be mapped to a spin chain, where exchange coefficients can be calculated from the shape of the trapping potential \cite{deuretz2,artem1}. Several works have detailed the static and dynamical properties of these models \cite{pu1,artem2,xiaoling2,deuretz_mdist,barfk3}. One interesting application of these tools is the investigation of the fundamental properties of quantum magnetism in models with a small number of particles \cite{amin2}. Recently, the magnetic correlations of a few-body strongly interacting system of fermions has been measured in the lab \cite{jochim3}. 

In this work, we focus on a yet unexplored regime of interactions in a one-dimensional strongly interacting Bose mixture. We show that, assuming an imbalance in the interaction between the two species it is possible to obtain radically different spatial correlations in a few-body system. Furthermore, we describe the dynamics of a mixture with imbalanced population in this interaction regime. We observe that the minority species can travel through the system as an effective spin wave packet in a certain interaction regime.

\section{Hamiltonian}\label{sec:1}

We consider the problem of a two-component Bose gas with population and interaction imbalance in an effective one-dimensional trapping geometry. We assume that the two different components consist of atoms of a same element, but in different hyperfine states. Therefore, the two bosonic species we take into account have the same mass $m$, but are defined by the states $\vert\uparrow\rangle$ or $\vert\downarrow\rangle$; the number of atoms in each species is respectively given by $N_\uparrow$ and $N_\downarrow$, with the total number of atoms written as $N=N_\uparrow+N_\downarrow$. We are interested in the spatial distributions and the dynamics of an imbalanced mixture where a minority species with $N_\downarrow$ atoms interacts with the remaining $N_\uparrow$ majority atoms. The general Hamiltonian for this system, considering contact interactions, is written as

\begin{eqnarray}\label{ham1}
H=\sum_{i=1}^{N} H_0(x_i)&+&g_{\uparrow\uparrow}\sum_{i<j}^{N_\uparrow}\delta(x_{\uparrow i}-x_{\uparrow j})\nonumber \\+ g_{\downarrow\downarrow} \sum_{i<j}^{N_\downarrow} \delta(x_{\downarrow i}-x_{\downarrow j}) &+& g_{\uparrow\downarrow} \sum_{i}^{N_\uparrow}\sum_{j}^{N_\downarrow} \delta(x_{\uparrow i}-x_{\downarrow j}),
\end{eqnarray}
where $H_0(x)=-\frac{\hbar^2}{2m}\frac{\partial}{\partial x}+V(x)$ is the single particle Hamiltonian with a trapping potential given by $V(x)$. In the following we assume that $\hbar=m=1$. We consider the general case of different intraspecies and interspecies interactions, that is $g_{\uparrow\uparrow}\neq g_{\downarrow\downarrow} \neq g_{\uparrow\downarrow}$. The particular case of $g_{\uparrow\uparrow} = g_{\downarrow\downarrow} = g_{\uparrow\downarrow}$ can be solved (for a homogeneous case) with the Bethe ansatz. Generally, the interactions between the different components can be tuned by means of Feshbach or confinement induced \cite{feshbach,olshanii} resonances. For a system with very strong interactions ($g\gg 1$), the Hamiltonian \eqref{ham1} can be mapped (up to linear order in $1/g$) to an effective spin chain. In the most general case of different interactions between the different components, this spin chain is described by \cite{xiaoling}
\begin{equation}\label{spinchain}
H_{S}=E_0 -\sum_{i=1}^{N-1}\alpha_i\left(\frac{1}{g_{\uparrow\uparrow}} P_{1,1}^{i,i+1}+\frac{1}{g_{\downarrow\downarrow}} P_{1,-1}^{i,i+1}+\frac{1}{g_{\uparrow\downarrow}} P_{1,0}^{i,i+1}\right),
\end{equation}
where $P_{S,M}$ denotes the projection operator on the two-particle eigenstates with total spin $S$ and magnetization $M$, while $E_0$ is the energy in the limit of infinite repulsion (see below). The exchange coefficients $\alpha_i$ are determined by the trapping geometry and can be thought of as being proportional to the spatial overlap of individual atoms in the trap \cite{deuretz2,pu1,artem1}. Different numerical methods to calculate these coefficients up to large values of $N$ are available and allow for the construction of effective hamiltonians with different trapping potentials \cite{conan,deuretz_mdist}. In the next sections, we will focus on a system trapped in the infinite square well, where all coefficients $\alpha_i$ have the same constant value. We point out that the results for the spatial distributions shown in the next section also hold for the case of a harmonic trap. Furthermore, we drop the term $E_0$, which only accounts for a constant energy factor.
Since we consider the bosonic case where no interaction between atoms is forbidden by the Pauli principle, the two-body scattering happens in the triplet channel
\begin{equation}\label{triplet}
\vert 1,1\rangle=\vert \uparrow\uparrow\rangle;\,\,\,\,
\vert 1,-1\rangle=\vert \downarrow\downarrow\rangle;\,\,\,\,
\vert 1,0\rangle=\frac{\vert\uparrow\downarrow\rangle+\vert\downarrow\uparrow\rangle}{\sqrt{2}},
\end{equation}
from which the projectors in Eq.\eqref{spinchain} can be built as 
\begin{eqnarray}
P_{1,1}=
  \begin{pmatrix}
    1 & 0 & 0 & 0 \\
    0 & 0 & 0 & 0 \\
    0 & 0 & 0 & 0 \\
    0 & 0 & 0 & 0 
  \end{pmatrix},\,
P_{1,-1}=
  \begin{pmatrix}
    0 & 0 & 0 & 0 \\
    0 & 0 & 0 & 0 \\
    0 & 0 & 0 & 0 \\
    0 & 0 & 0 & 1 
  \end{pmatrix},\
P_{1,0}=\frac{1}{2}
  \begin{pmatrix}
    0 & 0 & 0 & 0 \\
    0 & 1 & 1 & 0 \\
    0 & 1 & 1 & 0 \\
    0 & 0 & 0 & 0 
  \end{pmatrix}.
\end{eqnarray}
Following the same reasoning, spin chains for bosons or fermions with a higher number of internal components can be constructed \cite{decamp}. By expanding the projection operators in terms of the Pauli matrices \cite{xiaoling}, it is possible to reproduce a XXZ Hamiltonian (for $g_{\uparrow\uparrow}=g_{\downarrow\downarrow}\neq g_{\uparrow\downarrow}$) and the Heisenberg Hamiltonian with ferromagnetic correlations (for $g_{\uparrow\uparrow}=g_{\downarrow\downarrow}=g_{\uparrow\downarrow}$) \cite{guan2}.

\section{Spatial distributions}\label{sec:1}
We now focus on calculating the spatial distributions of a two-component bosonic system with population imbalance and arbitrary combinations of the interactions $g_{\uparrow\uparrow},g_{\downarrow\downarrow}$ and $g_{\uparrow\downarrow}$. We point out that, while the interactions may be different, the spin chain approach only remains valid as long as we have $\left(g_{\uparrow\uparrow},g_{\downarrow\downarrow},g_{\uparrow\downarrow}\right)\gg 1$. The spatial distributions of trapped few-body bosonic mixtures over wide interaction ranges has been studied, for instance, in \cite{miguel2,miguel,bruno}. Here, we calculate the densities for individual atoms in the trap and combine those results with the spin orderings obtained from the eigenstates of Eq. \eqref{spinchain}. The {\it spin densities} are thus defined as \cite{deuretz1}
\begin{equation}\label{spin densities}
\rho_{\uparrow,\downarrow}(x)=\sum_{i=1}^{N}\rho^{i}_{\uparrow,\downarrow}(x),
\end{equation}
where $\rho^i_{\uparrow,\downarrow}=m^{i}_{\uparrow,\downarrow}\rho^{i}(x)$ and $m^{i}_{\uparrow,\downarrow}$ is the magnetization probability at site $i$. The quantity $\rho^{i}(x)$ represent the individual distribution of atoms in the trap, and is given by 
\begin{equation}\label{onebody}
\rho^i(x)=N!\int_{\gamma} dx_1...dx_N \,\delta(x_i-x)|\Phi_0(x_1,...,x_i,...,x_N)|^2,
\end{equation}
where the integrals are performed in the region $x_1<...<x_i<...<x_N$. The wave function $\Phi_0(x_1,...,x_i,...,x_N)$ is a spinless fermion wave function, which is constructed as the Slater determinant of $N$ particles in the potential $V(x)$. In our case, those are simply the $N$ lowest eigenstates of the infinite square-well (the energy $E_0$, present in Eq.~\ref{spinchain}, is the sum of the energies of these orbitals). The spatial distribution obtained from $\Phi_0(x_1,...,x_N)$ thus coincides with the expected results from a Tonks-Girardeau gas, where the inifinite repulsion between the atoms mimics the Pauli exclusion principle \cite{girardeau}. Exploring the determinant form of $\Phi_0(x_1,...,x_N)$, Eq.~\eqref{onebody} can be written as \cite{deuretz1}

\begin{equation}\label{dets}
\rho^i(x)=\frac{\partial}{\partial x}\left( \sum_{j=0}^{N-1} c_{j}^{i} \frac{\partial^j}{\partial \lambda^j}\det \left[B(x)-1\lambda \right]\vert_{\lambda=0}\right),
\end{equation}
where $c_{j}^{i}=\frac{(-1)^{N-1}(N-j-1)!}{(i-1)!(N-j-i)!j!}$ and $B(x)$ is a matrix whose elements are given by the superpositions
$b_{mn}(x)=\int_{-\infty}^{x}dy\,\varphi_m(y)\varphi_n(y)$ of single-particle orbitals. Eq.~\eqref{dets} then allows us to avoid the multidimensional integration required for evaluating Eq.~\eqref{onebody}.

\begin{figure*}[h]
\centering
\includegraphics[width=1\textwidth]{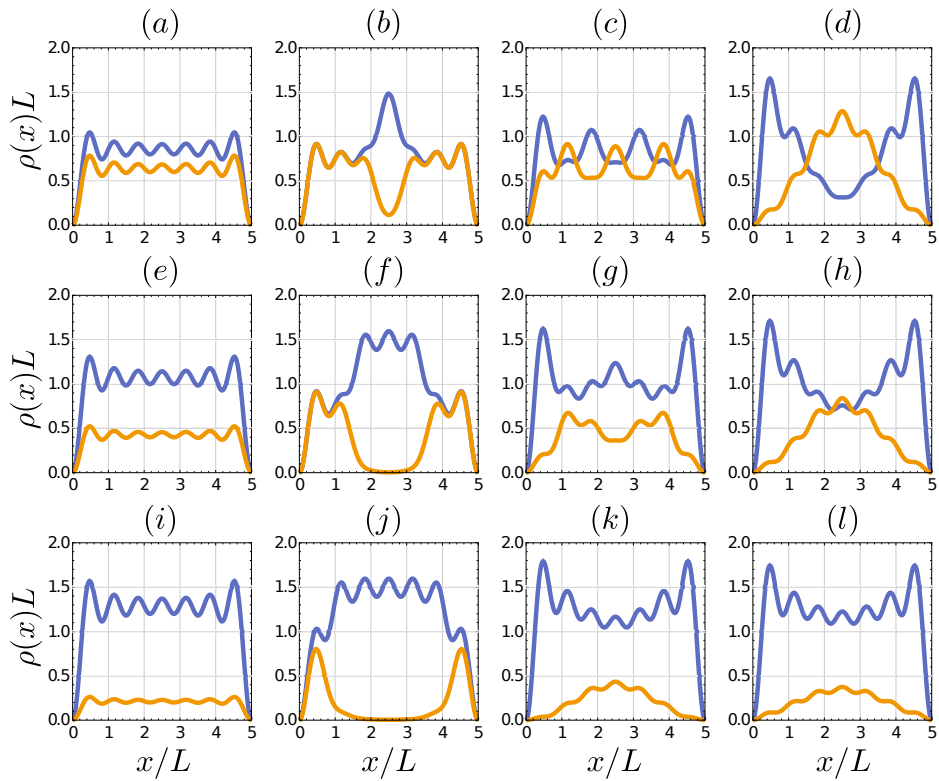}
\caption{Spin densities for a system with $N=7$ and $(a)$-$(d)$ $N_\uparrow =4,N_\downarrow=3$, $(e)$-$(h)$ $N_\uparrow=5,N_\downarrow=2$, $(i)$-$(l)$ $N_\uparrow=6,N_\downarrow=1$. The blue curves show the densities $\rho_{\uparrow}(x)$, while the yellow curves show $\rho_{\downarrow}(x)$. Each column describes an interaction regime (see text): $(a),(e),(i)$ HFM ($g_{\uparrow\uparrow}=g_{\downarrow\downarrow}=g_{\uparrow\downarrow}$); $(b),(f),(j)$ IFM ($g_{\uparrow\uparrow}=g_{\downarrow\downarrow}<g_{\uparrow\downarrow}$); $(c),(g),(k)$ AFM ($g_{\uparrow\uparrow}=g_{\downarrow\downarrow}>g_{\uparrow\downarrow}$) and $(d),(h),(l)$ IMB ($g_{\downarrow\downarrow}<g_{\uparrow\downarrow}<g_{\uparrow\uparrow}$). The density for each component is normalized to the number of particles.}
\label{fig:1}
\end{figure*}

In Fig.~\ref{fig:1} we show the results for the distributions calculated with Eq.~\eqref{spin densities} for $N=7$ atoms and different population imbalances: $N_\uparrow =4,N_\downarrow=3$, $N_\uparrow=5,N_\downarrow=2$, and $N_\uparrow=6,N_\downarrow=1$. We also consider different interaction regimes, which we define as: the Heisenberg ferromagnetic (HFM) regime ($g_{\uparrow\uparrow}=g_{\downarrow\downarrow}=g_{\uparrow\downarrow}$), the Ising ferromagnetic (IFM) regime ($g_{\uparrow\uparrow}=g_{\downarrow\downarrow}<g_{\uparrow\downarrow}$), the antifferomagnetic regime (AFM) ($g_{\uparrow\uparrow}=g_{\downarrow\downarrow}>g_{\uparrow\downarrow}$) and the completely imbalanced (IMB) regime ($g_{\downarrow\downarrow}<g_{\uparrow\downarrow}<g_{\uparrow\uparrow}$). Each of these cases corresponds to a column in Fig.~\ref{fig:1}: in the column defined by panels $(a),(e)$ and $(i)$, we observe the results for the HFM regime of identical interactions, where the distributions are the same, only scaled to the number of particles in the specific component. The profiles show ferromagnetic correlations of the Heisenberg type, which are expected for isospin bosons \cite{guan2}. Column $(b),(f),(j)$ show the IFM case where the intraspecies interaction is smaller than the interspecies interactions. This regime is characterized by a phase separated distribution, typical of an Ising ferromagnet. In column $(c),(g),(k)$ we present the AFM case where the intraspecies interactions are larger than the interspecies interactions. Here, we observe that antiferromagnetic correlations arise, being particularly visible in the slightly imbalanced case of panel $(c)$. Finally, we consider in column $(d),(h),(l)$ a completely imbalanced regime (IMB), where $g_{\downarrow\downarrow}<g_{\uparrow\downarrow}<g_{\uparrow\uparrow}$. Here, we see that we again obtain phase-separated profiles, excpet that now the minority atoms sit in the center of the trap. This happens because the minority intraspecies interactions are smaller and the net result is an effectively attractive regime for this component. We note that these distributions resemble the $XY$ phase described in a Bose-Fermi mixture \cite{deuretz3}. The evident similarity between the last two panels $(k)$ and $(l)$ results from the fact that in both cases there is only one minority particle. Therefore, the intraspecies interaction in this case has no effect on the minority spin density.

\section{Dynamics}\label{sec:2}
\begin{figure*}[h]
\centering
\includegraphics[width=0.75\textwidth]{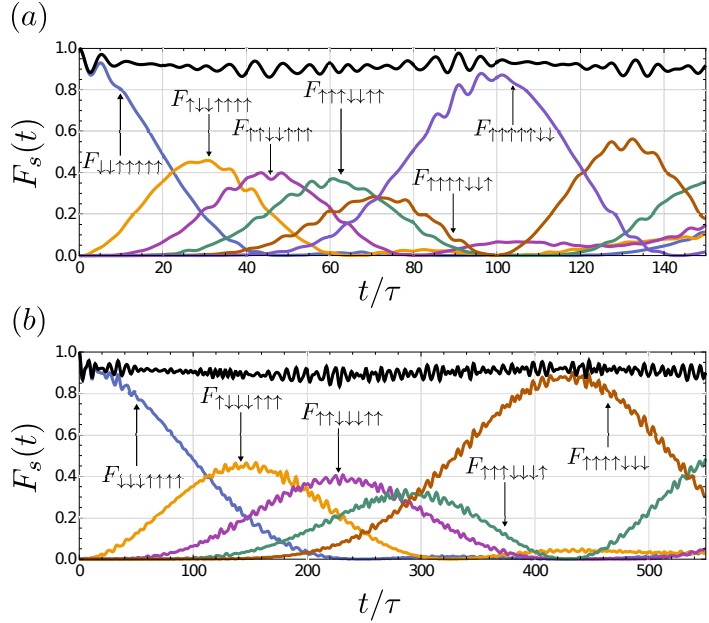}
\caption{Time evolution of the overlap probabilities $F_s(t)$ for the population imbalances $(a)$ $N_\uparrow=5,N_\downarrow=2$ and $(b)$ $N_\uparrow =4,N_\downarrow=3$, in the regime where $g_{\downarrow\downarrow}=g_{\uparrow\uparrow}/5=g_{\uparrow\downarrow}/3$. Each curve shows the projection over a state with a different spin ordering, as indicated by the arrows. The black curves show the sum of the projections over all states $\vert s \rangle$.}
\label{fig:2}  
\end{figure*}

We now turn to the dynamics of the system in the case of population and interaction imbalance. We choose to analyze the two cases of $N_\uparrow=5,N_\downarrow=2$ and $N_\uparrow =4,N_\downarrow=3$, with interactions strength ratios set as $g_{\uparrow\downarrow}=3g_{\downarrow\downarrow}$ and $g_{\uparrow\uparrow}=5g_{\downarrow\downarrow}$. The case of a single minority spin in different interaction regimes and trapping potentials has been studied in \cite{artem2,loft1}. We will show that, in the imbalanced regime, non-trivial effects for the dynamics of the minority species can arise. The system is initialized in the state where the minority species is placed at the left side, that is $\vert \psi(0)\rangle=\vert \downarrow_1 ... \downarrow_{N_\downarrow}\uparrow_1 ... \uparrow_{N\uparrow}\rangle$. We disregard any external excitations to the system, such that the dynamics can be described completely by Eq.~\eqref{spinchain}. The time unit $\tau=g_{\downarrow\downarrow}\pi/\alpha$ is set by the energy scale of the trap through the geometrical coefficient $\alpha$. We calculate the time-dependent overlap of the wave function with a state $\vert s \rangle$ defined by a given spin ordering as
\begin{equation}
F_{s}(t)=\vert \langle s \vert e^{-i H_S t}\vert \psi(0)\rangle \vert^2.
\end{equation} 
In Fig.~\ref{fig:2} we show the results for the cases of $(a)$ $N_\uparrow=5,N_\downarrow=2$ and $(b)$ $N_\uparrow =4,N_\downarrow=3$. We focus only in the cases where $\vert s \rangle$ denotes a state where the minority spins are ``bunched up" the system, e.g. $\vert \downarrow\downarrow\uparrow\uparrow\uparrow\uparrow\uparrow\rangle$, $\vert \uparrow\downarrow\downarrow\uparrow\uparrow\uparrow\uparrow\rangle$, and so on.

We clearly observe in both imbalanced cases that the minority spins, initialized in the left, travel through the system together, with higher amplitudes for the projections over the states with the minority at the edges. We also show the sum of the projections over all states $\vert s \rangle$, which nearly amounts to unity at all times, indicating that other spin orderings have little influence on the dynamics. The physical interpretations for this phenomenon comes from the fact that the minority intraspecies interaction is weaker than the remaining two. Additionally, the repulsion between the majority atoms dominates. This leads to a regime where the minority atoms behave as if under the effect of an attractive force. While this combination of strongly imbalanced interactions may be hard to achieve in practice, it is possible to devise a scenario where the majority strongly repulsive bosons are replaced by a Fermi gas. Bose-Fermi mixures have been studied under the same formalism applied here \cite{deuretz3}, and could present similar dynamical effects.

\section{Conclusions}\label{conc}
We have studied the effects of imbalance in the spatial correlations and dynamics of a strongly interacting two-component Bose gas. The results for spin densities reproduce the expected behavior in the known regimes of ferromagnetic and antiferromagnetic correlations. Furthermore, the completely imbalanced case can present non-trivial spatial distributions that are not found in other regimes. The dynamics of the system is also affected by the population and interaction imbalance. We demonstrated that, given a certain choice of interaction parameters, a minority population initialized at the edge of the system can move around as if bound by an effectively attractive force. Our work suggests that set ups where the interactions can be tuned between different species may be ideal for studying the transfer of spin packets. Such a realization could provide knowledge of quantum transport in spin chains beyond the case of a single impurity. By generalizing the spin chain approach to a higher number of internal components, it is also possible to consider cases of mixed multicomponent systems with different interactions in several scattering channels.

\begin{acknowledgements}
The authors thank Artem G. Volosniev for feedback on the results. The following agencies - Conselho Nacional de Desenvolvimento Cient{\'i}fico e Tecnol{\'o}gico (CNPq), the Danish Council for Independent Research DFF Natural Sciences and the DFF Sapere Aude program - are gratefully acknowledged for financial support. 
\end{acknowledgements}

\end{document}